# Laser Processing For 3D Junctionless Transistor Fabrication

D. Bosch[1,2], P. Acosta Alba[1], S. Kerdiles[1], V. Benevent[1], C. Perrot[3], J. Lassarre[1], J. Richy[1], J. Lacord[1], B. Sklenard[1], L. Brunet[1], P. Batude[1], C. Fenouillet-Béranger[1], D. Lattard[1], J. P. Colinge[1], F. Balestra[2], F. Andrieu[1]

[1]CEA-LETI, Univ. Grenoble Alpes, 17 rue des Martyrs, 38054 France ; email : daphnee.bosch@cea.fr; [2]Univ. Grenoble Alpes, CNRS, Grenoble INP, IMEP-LAHC, F-38000 France, [3]STMicroelectronics, 850 rue Jean Monnet, F38926 Crolles.

***Abstract*—To take fully advantage of Junctionless transistor (JLT) low-cost and low-temperature features we investigate a 475°C process to create onto a wafer a thin poly-Si layer on insulator. We fabricated a 13nm doped (Phosphorous, $10^{19}$ at/cm$^3$) poly-silicon film featuring excellent roughness values ($R_{max}$= 1.6nm and RMS=0.2nm). Guidelines for grain size optimization using nanosecond (ns) laser annealing are given.**

## I. Introduction

3D-monolithic integration, also named 3D sequential integration, consists in stacking active layers on top of each other in a sequential manner onto a given wafer substrate (Fig. 1). Direct wafer bonding can be used to transfer the top silicon layer of a Silicon-On-Insulator (SOI) substrate onto a processed bottom wafer for the channel formation. This approach requires the use of an SOI wafer and is very challenging in terms of thermal budget to avoid bottom tier degradation. In fact, a maximal thermal budget of 500°C during 2 hours has been establish to preserve the integrity of the bottom tier [1-3].

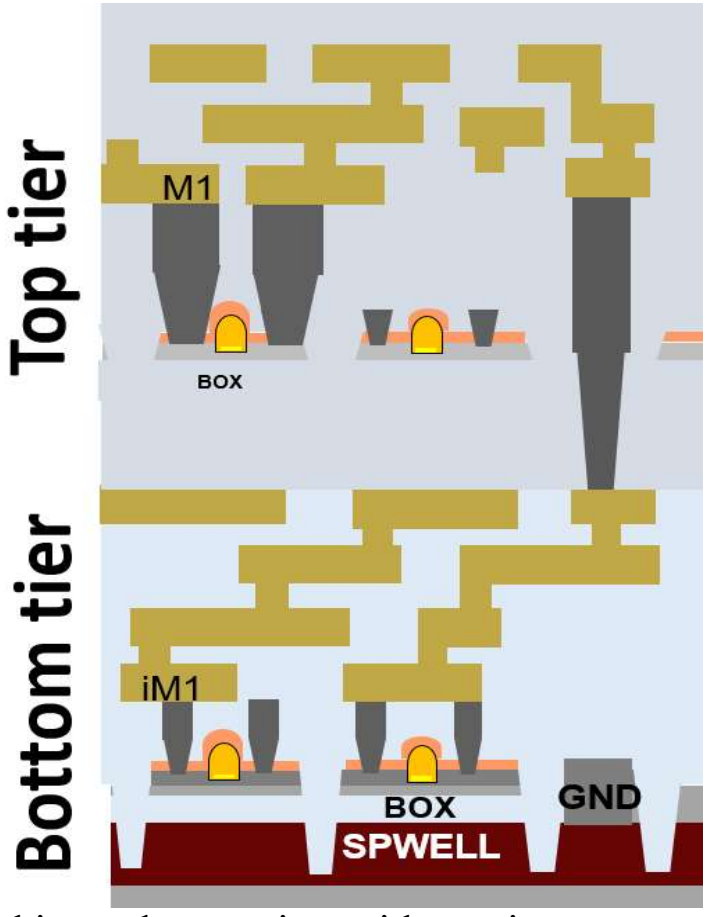

**Fig.1:** 3D monolithic stack overwiew with two interconnected tiers.

Thus, a good candidate for top tier devices is the junction-less transistor (JLT), for which the uniform device doping avoids high-temperature Source-Drain (SD) formation [4]. Furthermore, the channel material can be directly deposited on a bottom tier without damages [5-6]. However, this integration scheme suffers from poly-silicon roughness obtained after processing. For instance, typical RMS values are 0.7nm [5]-1.2nm [6] (Green Nanosecond Laser Crystallization (GNS-LC) + Chemical Mechanical Polishing (CMP)) or 0.6nm [7] (HPA trimming).

To take benefits from JLT low cost feature, this work is focused on the following process: low-temperature amorphous silicon (a-Si) deposition followed by nanosecond laser annealing and CMP (Fig. 2). The specifications are the following: a maximum thermal budget of 500°C, 2hours, and a 12-15nm thick a-Si layer doped at 1e$^{19}$ at/cm$^3$ for future JL device formation with a ≤0.5nm RMS variation to lower device variability.

## II. PROCESS FLOW

### A. *475°C Low Pressure Chemical Vapor Deposition*

A 35 nm thick a-Si layer is deposited at 475°C on an oxidized blanket bulk wafer. *Ex-situ* doping (using ion implantation) and in-situ doping have been compared. For *in-situ* doping deposition, a 90sccm phosphorus flow is used in order to achieve a $10^{19}$ at/cm$^3$ doping concentration [8]. For *ex-situ* doping, the implantation conditions have been simulated by Kinetic Monte-Carlo TCAD (Silvaco).

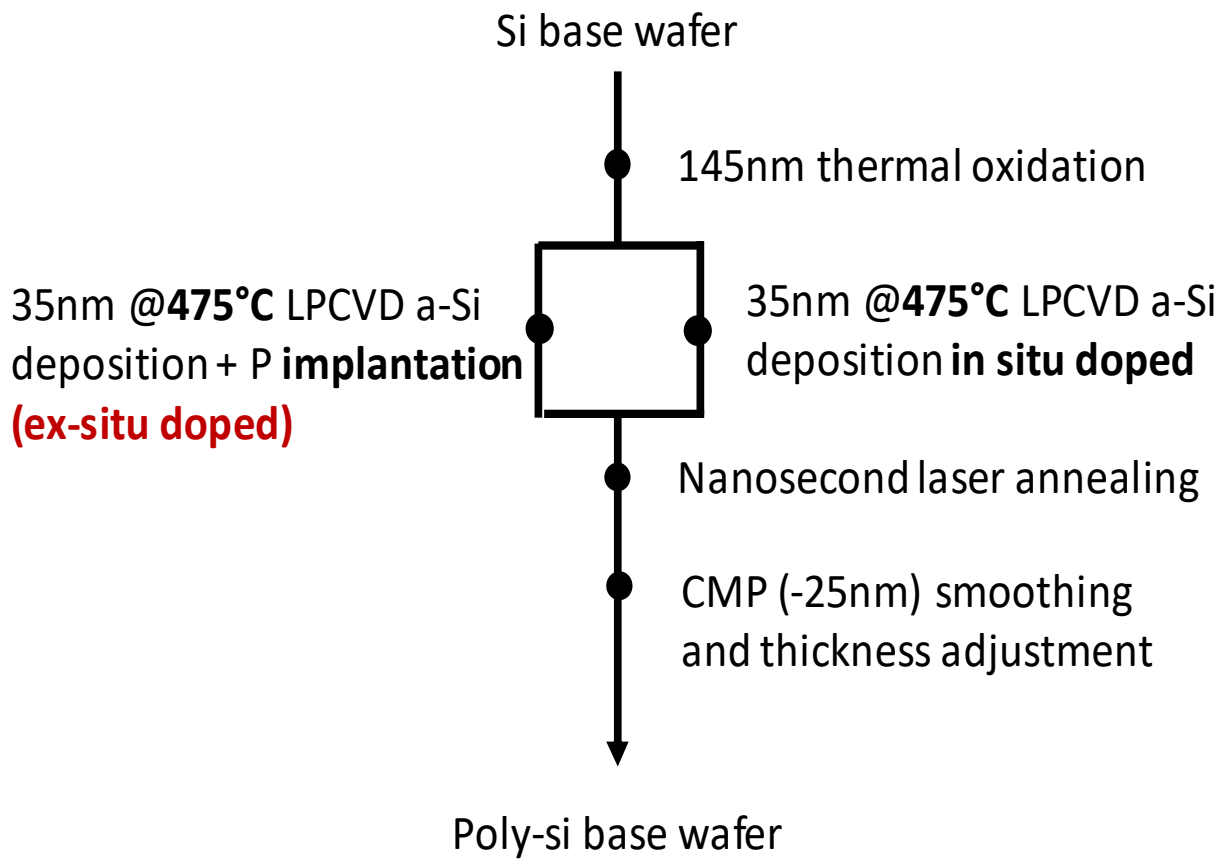

**Fig.2:** Integration scheme. *In-situ* and *ex-situ* doping are compared.

### B. *UV Nanosecond (ns) Laser Annealing (UV-NLA)*

Thanks to low depth penetration, UV-NLA is suitable for crystallizing a top a-Si layer while preserving the integrity of the bottom tier [9]. For these reasons, an excimer laser (wavelength: 308nm, optimized pulse duration 160ns [10]) is used to activate the dopants and recrystallize the a-Si layer.

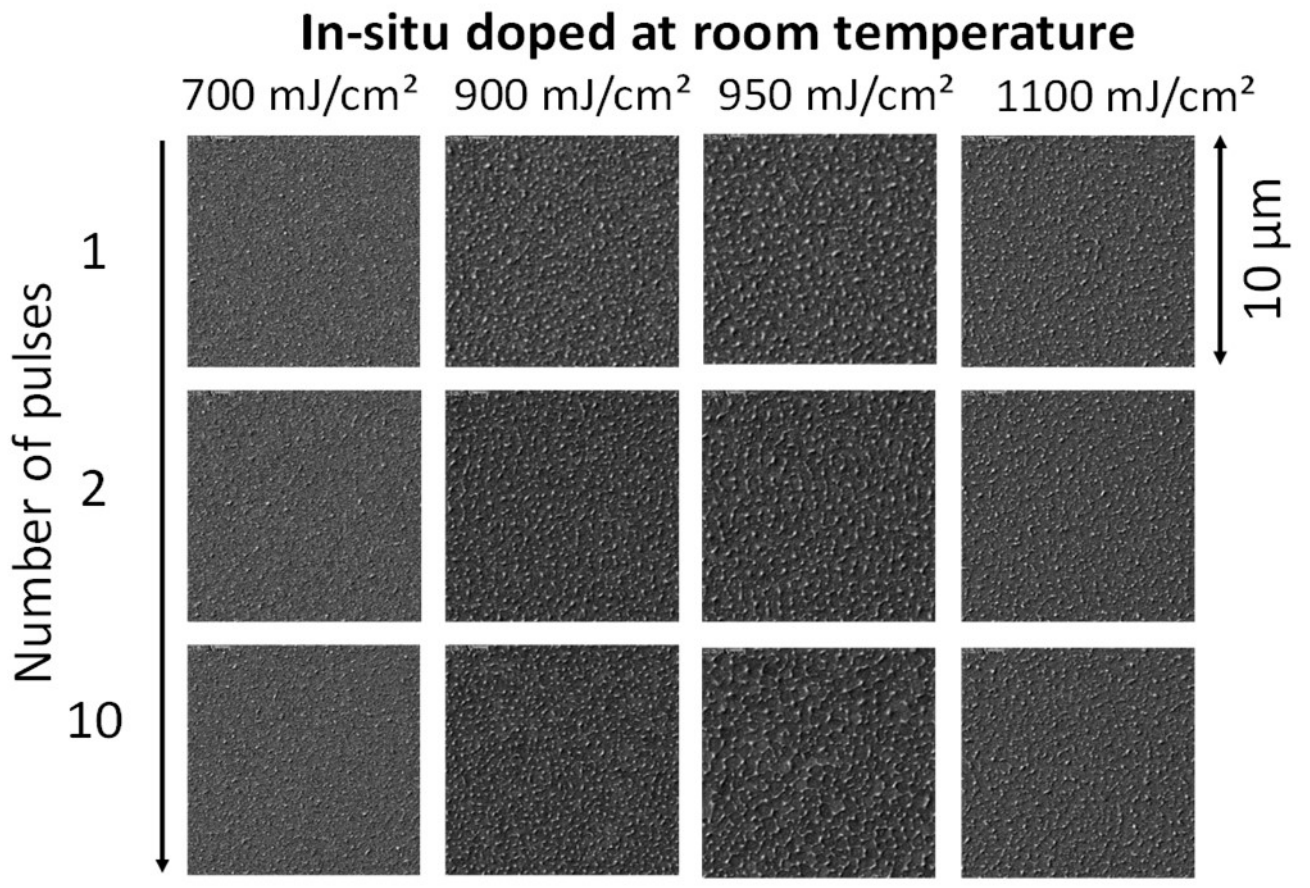

**Fig.3:** Scanning Electron Microscopy (SEM) images showing the impact of cumulated pulses for different laser energy density values.

Based on Time Resolved Reflectometry (TRR) analysis (energy density screening from 0.3 to 1.4 J/cm²) three regimes can be identified (Figs 3-6). In fact, *in-situ* reflectometry monitoring allows us to detect film melting (Reflectivity decreases).

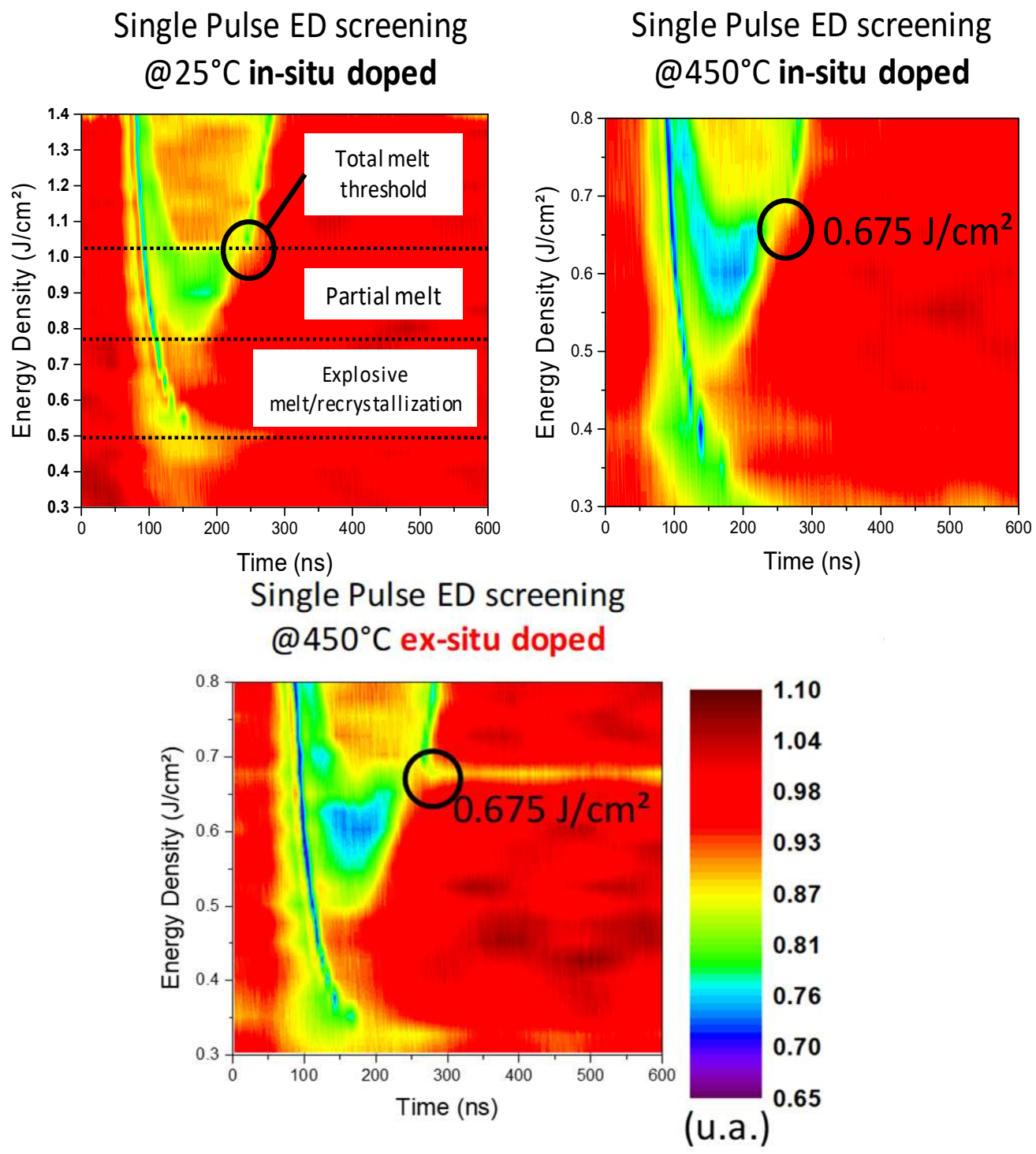

**Fig.4:** Time Resolved reflectometry: energy density screening. Three regimes are observed: explosive melt/spontaneous recrystallization, partial melt and total melt. No difference observed between *in-* and *ex-situ* doping.

For low energy density (<0.775 J/cm²), no melting is detected though TRR. An explosive melt followed by a spontaneous recrystallization can occur for such energy densities (see Fig.4). For intermediate energy densities (preferred processing window), a part of the a-Si film melts and recrystallizes, forming grains and lowering the film resistivity. For higher energy density (> 1.025J/cm²), the layer is entirely melted before the end of the laser pulse resulting in an amorphous film.

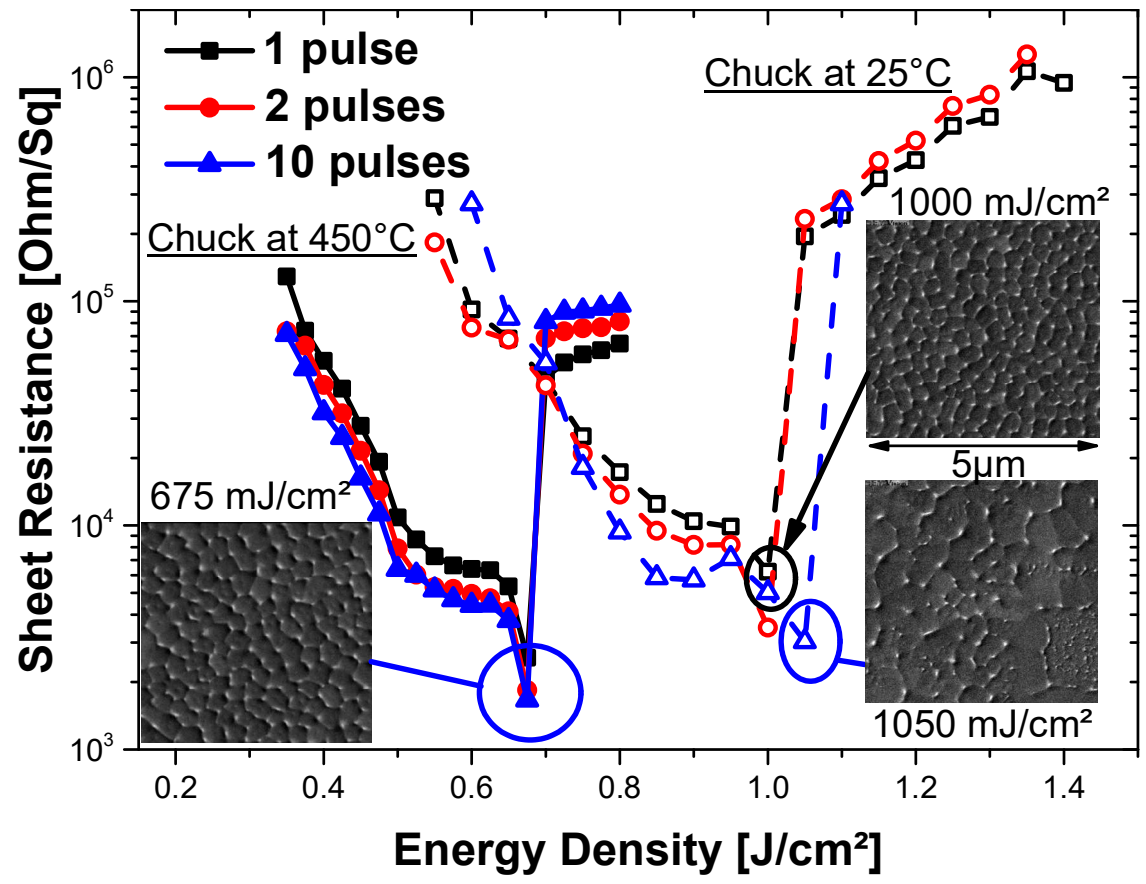

**Fig.5:** Impact of chuck temperature and number of pulses on sheet resistance for in-situ doped Si

Moreover, maintaining the stage temperature at 450°C reduces the thermal gradient undergone by the film. It could slow down the cooling, yielding larger grain and thus lower resistivity [11]. For instance, at the melt threshold, melting time is 156ns (chuck at 450°C) compared to 130ns at 25°C. In addition, a 2nd laser annealing (two pulses) can be used to preferentially melt the smaller grains. These coalesce with grains leading to an overall grain size increase [12].

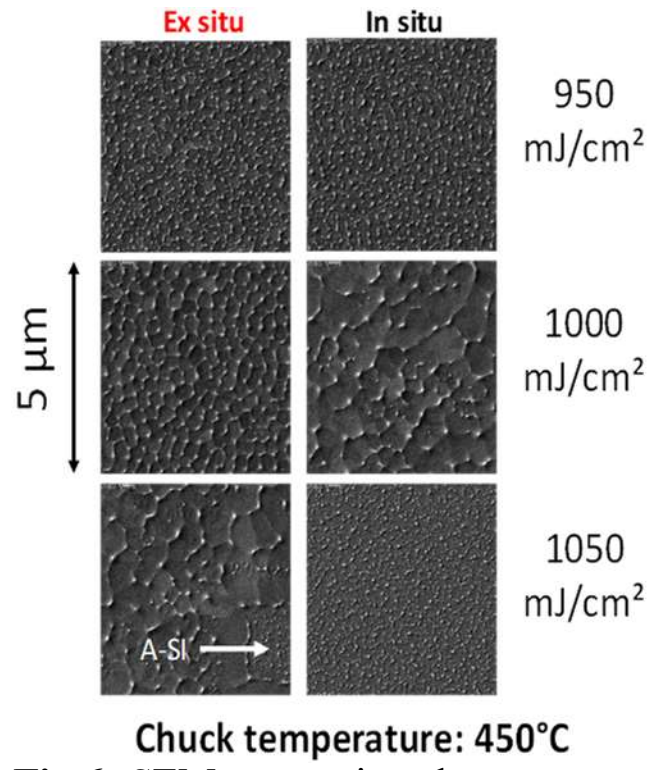

**Fig.6:** SEM comparison between *in-* and *ex-situ* doping. No significant differences is seen.

| | Stage temperature | Number of pulses | Label | Energy (J/cm²) |
|---|---|---|---|---|
| In situ doped | 450 °C | 4 | $I_1$ | 0.625 |
| | | | $I_2$ | 0.675 |
| Ex situ doped | 450°C | 4 | $E_1$ | 0.625 |
| | | | $E_2$ | 0.675 |

**Fig.7:** Selected conditions to maximize grain size and obtain a low resistivity and lower processing time.

As far as the comparison of *in-* and *ex-situ* doping is concerned (Figs 6-9), no significant difference can be observed in terms of energy response and grain size. To probe further, two energy density conditions are studied (0.625J/cm² for $I_1$, $E_1$, and 0.675J/cm² for $I_2$, $E_2$, as defined in Fig.7) with a stage temperature of 450°C using up to four cumulative pulses.

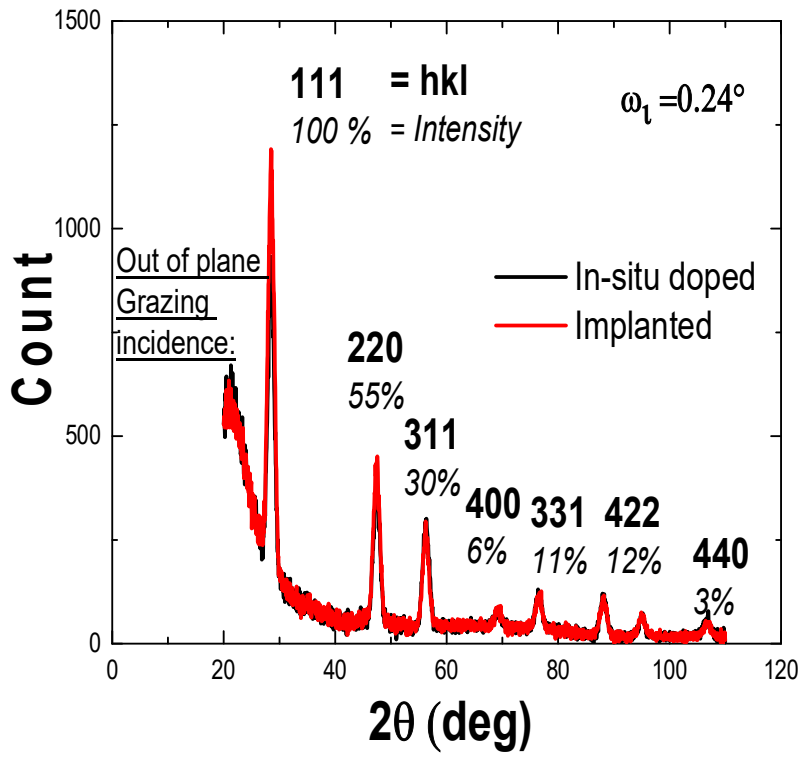

**Fig.8:** XRD out-of plane grazing incidence patterns. Miller index and intensity are taken from [13]

Patterns from grazing incidence X-ray diffraction in in-plane and out-off-plane geometry (Fig. 8) correspond to poly-Si with no texture, indicating no preferential direction regrowth during annealing.

### *C. Chemical mechanical polishing (CMP)*

| Extracted from 5µm*5µm (AFM) | | In-situ doped | | Ex-situ doped | |
|---|---|---|---|---|---|
| | | $I_1$ | $I_2$ | $E_1$ | $E_2$ |
| Before CMP | $R_{MAX}$ (nm) | 50 | 63 | 49 | 71 |
| | $R_Q$(nm) | 9 | 9.6 | 7 | 10 |
| After CMP | $R_{MAX}$ (nm) | 4 | **2.3** | **1.6** | **1.7** |
| | $R_Q$(nm) | 0.25 | **0.29** | **0.2** | **0.2** |

**Fig.9:** Roughness measured by Atomic Force Microscopy (AFM) on 5x5µm² scan before and after CMP.

To adjust the thickness and to lower the roughness, 25nm silicon is removed by CMP. For instance, for the $E_1$ condition, the peak-to-valley thickness variation ($R_{MAX}$) is reduced from 49 nm (before CMP) to 1.6 nm (after CMP) and RMS ($R_Q$) from 7nm to 0.2 nm (Fig. 9).

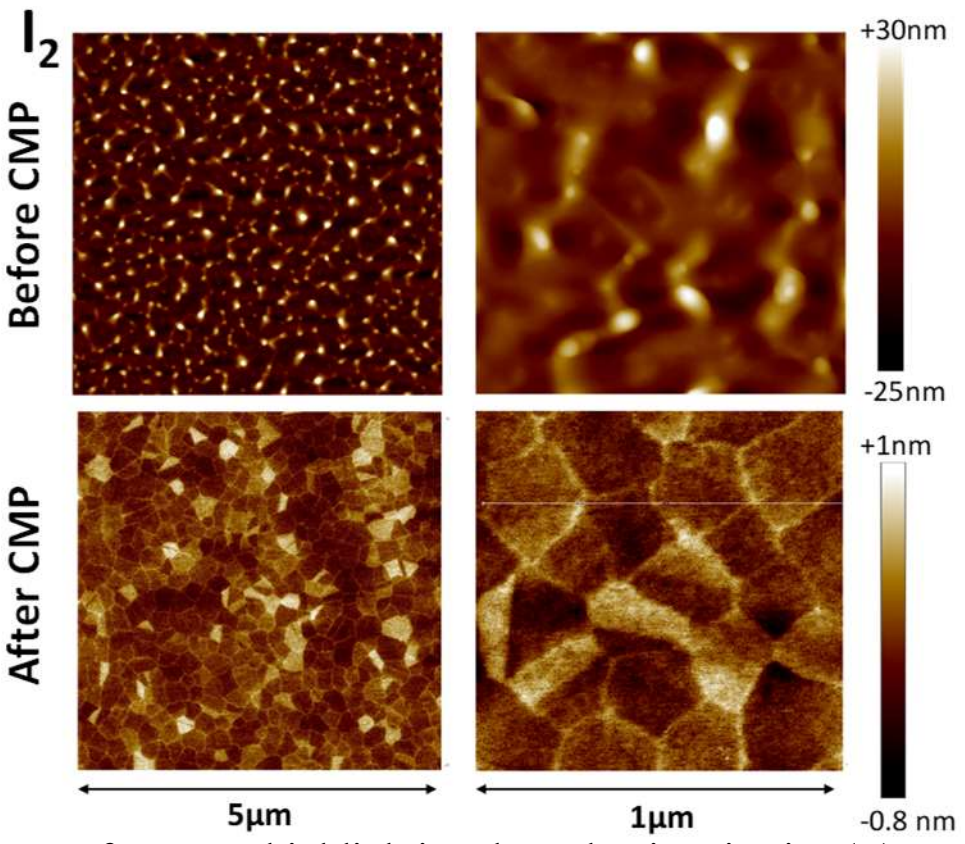

**Fig.10:** AFM after CMP highlighting the poly-si grain size ($I_2$)

Furthermore, Atomic Force Microscopy measurements (Fig. 10) evidence that grain size is unchanged after CMP, highlighting that the film morphology is homogeneous within the depth.

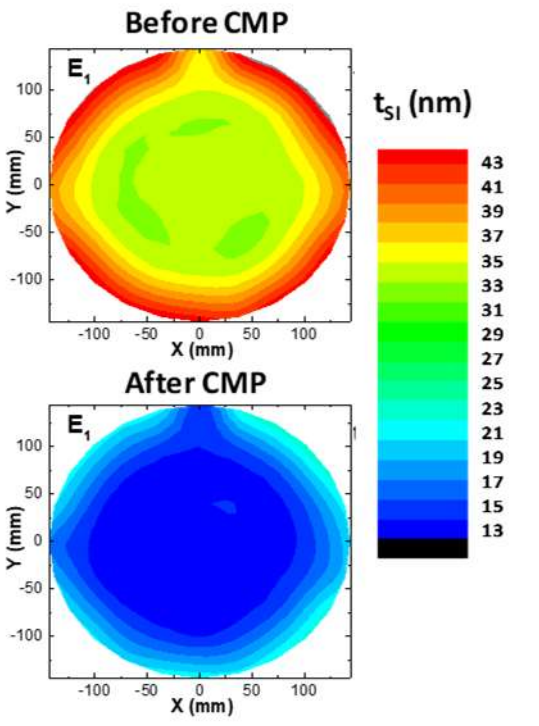

**Fig.11:** Silicon thickness before/after CMP ($E_1$).

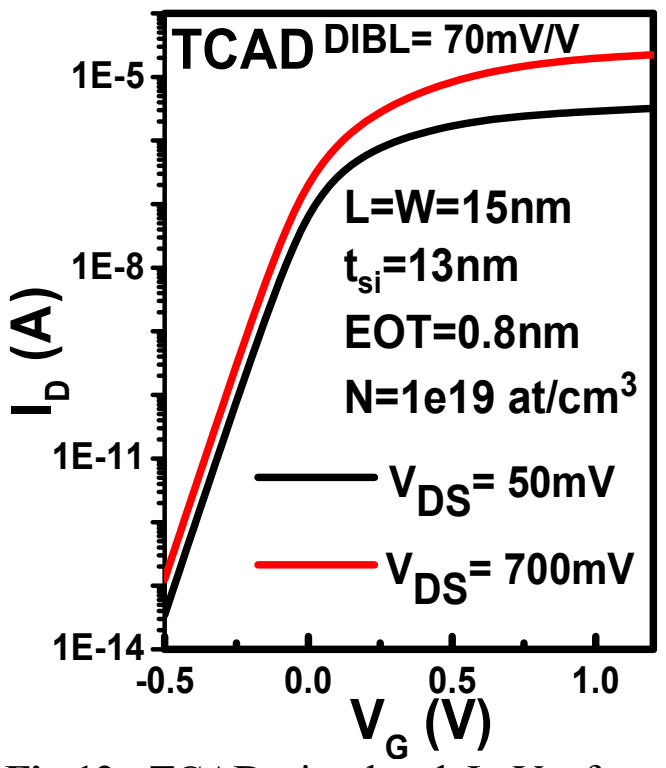

**Fig.12:** TCAD simulated $I_D$-$V_G$ for a junctionless transistor featuring 13nm $10^{19}$ doped channel.

Finally, a 13 nm average thickness is achieved (Fig. 11), which is appropriate for 15nm gate length and width JL transistor (equivalent oxide thickness of =0.8nm and DIBL= 70mV/V) (Fig. 12).

## III. Conclusion

We provide guidelines, to form thin doped poly-Si film on insulator suitable for JLT 3D monolithic low temperature and low cost integration (475°C). When considering processing time and thus cost, *in-situ* doping, 450°C stage temperature and low laser pulses number are preferred for optimized recrystallization.

## ACKNOWLEDGMENTS:

This work was supported by French Public Authorities through LabEx Minos ANR-10-LABX-55-01. We would like to thanks SCREEN company and its French subsidiary LASSE for their support in operating and maintaining the LT-3100 laser annealing platform.